# The Orbit of Asteroid 1994 PC1
# (July 27, 2011)

Sophia Jane Balkoski[1], Laksh Bhasin[1], and Milly KeQi Wang[1]

[1]Summer Science Program, Santa Barbara, CA USA

**ABSTRACT:**

Near-Earth Asteroids can be hazardous to the Earth, due to their orbital characteristics and proximity to inner Solar System planets. Using three sets of CCD images collected in June and July 2011, the orbital elements of asteroid 1994 PC1 were determined at solar opposition. The body's specific right ascension and declination were calculated through least squares plate reduction (taking parallax into account) and compared to those of the Jet Propulsion Laboratory. These data were then used to find 1994 PC1's orbital elements, as well as any statistical uncertainty. This research points to an eccentricity of 0.3272 (+0.00009 or –0.0005), a semi major axis of 1.3401 AU (+0.0001 AU or -0.0004 AU), an inclination angle of 33.30$^o$ (+0.05$^o$ or -0.01$^o$), a longitude of the ascending node of 118.01$^o$ (+0.09$^o$ or -0.02$^o$), an argument of perihelion of 47.14$^o$ (+0.005$^o$ or -0.08$^o$), and a time of last perihelion of JD 2455642.88 (+0.007 or -0.03).

## I. INTRODUCTION

NEAR-EARTH Asteroids are potentially hazardous celestial bodies with a perihelion distance no greater than ≈ 1.3 Astronomical Units. These objects can have elliptical orbits with high eccentricity and, due to their relative abundance in the Solar System, are liable to collide with the Earth. For example, the Asteroid 99942 Apophis – which was discovered in 2004 at the Kitt Peak National Observatory – may hit the Earth as soon as 2029 with more than twice the destructive force of the Krakatoa volcanic eruption in 1883.[1] This paper studies the orbital elements of near-earth asteroid 1994 PC1 (also known as Asteroid 7482) using data collected over the course of June and July 2011 using a 14-inch Meade telescope located in Santa Barbara, California. Though this report concerns but one of many potentially dangerous asteroids, it nonetheless illustrates the importance of continued astronomy research; the data gathered can be used both for the advancement of science and for the sake of public safety.

The orbital elements were determined through the Method of Gauss (an iterative process using differential calculus), Kepler's laws, and Newtonian gravity. A simulation of both the asteroid and the Solar System was carried out over a period of 100 years. This showed that the distance of closest approach between 1994 PC1 and Earth would be ≈ 0.0494 AU on January 21, 2056 (ignoring the accumulation of uncertainties over time).

## II. PROCEDURE OF OBTAINING ASTEROID DATA

### 1) Observing Specifications

A Meade LX200-ACF telescope with a 14-inch aperture was used to observe the asteroid. The focal length of this telescope is 3556mm f/10 (14"). Its field of view is 21 inches x 16.8 inches.[2] The telescope was located at latitude 34° 26' 43.89" N and longitude 119° 39' 41.97" W. The elevation was 150 meters. The telescope was placed on an equatorial mount. The star-mapping program TheSkyX was linked to the telescope and used to locate the asteroid.

An SBIG STL-1301E Charge-coupled Device (CCD) camera is attached to the telescope. The pixel array is 1280 x 1024 pixels or 20.5 x 16.4 mm. There are 1.3 million pixels in total. The pixel size is 16 x 16 microns. 2 x 2 binning were used for the asteroid images. The CCD camera must be cooled down to -15°C before obtaining images.[3]

The software CCDsoft was used to focus the stars and take images. In order to track the asteroid's movement across the sky, several series of images were taken with fifteen-minute intervals in between. A feature in CCDsoft was used to align and blink the sets of images to help determine the location of the asteroid.

### 2) Observing Methods

Prior to each observation, the rough right ascension (RA) and declination (Dec) of the asteroid were found using the JPL Horizons website. These were known as "pointing coordinates." Several stars in the vicinity of the asteroid (as well as their RA and Dec) were recorded beforehand. The hour angle of the asteroid was calculated using its RA and the Local Sidereal Time. This gave a rough estimate of the position of the asteroid in the sky. All of this information, including a star chart, was recorded in each researcher's observation notebooks.

Due to the potential mechanical error of the telescope, the telescope had to be synced to a sync star at the beginning of each observation. Any bright and well-documented star would qualify as a sync star. Since the telescope was placed on an equatorial mount, it would flip while tracking the asteroid if the asteroid were to cross the meridian. Once the telescope flips, it needs to be re-synced to a new star. East of the local meridian, Vega served as an adequate sync star and, west of the local meridian, Arcturus was used.

The RA and Dec of the asteroid were entered into TheSkyX and the telescope was slewed to that portion of the sky. Before taking any images, the stars were focused using CCDsoft. Once the focus was adequate, a series of five or seven images with 37 to 45 second exposures were taken. Three series of images were taken during each observation night; these were then aligned and blinked to identify the movement of the asteroid. Any observations concerning ambient conditions (i.e. astronomical seeing) were recorded in the observing notebooks.

### 3) Technical Problems and Solutions

JPL Horizons contains both the apparent RA and Dec of the asteroid as well as the astrometric RA and Dec. After a few observations, it was determined that the apparent RA and Dec resulted in the asteroid appearing at the edges of the CCD image. As a result, the astrometric RA and Dec were used for subsequent observations to obtain an image with a centered asteroid.

### 4) Techniques for Analyzing Data

The RA and Dec of the asteroid can be calculated using the RA and Dec of surrounding reference stars. The centroids of six to eight surrounding reference stars were found using a Gaussian fit on MaxIm DL. The centroids for each star were then grouped with that star's recorded RA and Dec. Using this information, it was possible to determine the RA and Dec of the asteroid through the Least-Squares Plate Reduction (LSPR) method. The covariance and correlation (r) values were also calculated. The plate solution equations (required for LSPR) for the CCD images are as follows:

$$\alpha = b_1 + a_{11}x + a_{12}y \quad (1)$$
$$\delta = b_2 + a_{21}x + a_{22}y \quad (2)$$

Where $\alpha, \delta$ are the RA and Dec of an object in the image, $(x, y)$ are the x and y positions of the object in the CCD image in units of pixels, and the other terms are the six plate solution coefficients.

After obtaining calculated RA and Dec values of the asteroid for three observations, the Method of Gauss was used to find the vector orbital elements (position and velocity vector). The position vector ($\vec{r}$) can be found using the equation:

$$\rho\hat{\rho} = \vec{r} + \vec{R} \quad (3)$$

Where $\rho$ is the distance from the observer to the asteroid (the "range"), $\hat{\rho}$ is the unit vector pointing from the observer towards the asteroid, $\vec{r}$ is the vector from the center of the Sun to the center of the asteroid, and $\vec{R}$ is the vector from the observer to the center of the Sun. The f and g series can be used to approximate the range vector over several iterations of the Method of Gauss.

Using the vector orbital elements, the classical orbital elements (semi-major axis, eccentricity, angle of inclination, longitude of ascending node, argument of perihelion, and time of last perihelion) were calculated.

The vector orbital elements ($\vec{r}$ and $\dot{\vec{r}}$) were then corrected through differential corrections. These corrections involved computing ephemeris values for RA and Dec (based on calculated classical elements) and comparing them to our three pairs of observed values ($\alpha_1, \alpha_2, \alpha_3, \delta_1, \delta_2, \delta_3$). This would give $\Delta\alpha_1, \Delta\alpha_2, \Delta\alpha_3, \Delta\delta_1, \Delta\delta_2$, and $\Delta\delta_3$.

Partial derivatives of each observed RA and Dec (with respect to the x, y, and z components of position and velocity) were also required. These were obtained through numerical differentiation. For example, suppose $\vec{r} = \{x_0, y_0, z_0\}$ and $\dot{\vec{r}} = \{\dot{x}_0, \dot{y}_0, \dot{z}_0\}$. This means:

$$\frac{\partial \alpha_1}{\partial x_0} \approx \frac{\alpha(x_0 + \Delta x_0, y_0, z_0, \dot{x}_0, \dot{y}_0, \dot{z}_0) - \alpha(x_0, y_0, z_0, \dot{x}_0, \dot{y}_0, \dot{z}_0)}{\Delta x_0}$$

Where $\Delta x_0$ is any arbitrary small number that will still produce a measurable change (e.g. $10^{-6}$) and $\alpha(x_0, y_0, z_0, \dot{x}_0, \dot{y}_0, \dot{z}_0)$ gives RA as a function of position and velocity.

Based on the partial derivatives and the ephemeris-generated residuals in RA and Dec, the following system of linear equations can be used to perform corrections:

$$\begin{bmatrix}\Delta\alpha_1\\\Delta\alpha_2\\\Delta\alpha_3\\\Delta\delta_1\\\Delta\delta_2\\\Delta\delta_3\end{bmatrix} = \begin{bmatrix}\frac{\partial\alpha_1}{\partial x_0} & \frac{\partial\alpha_1}{\partial y_0} & \frac{\partial\alpha_1}{\partial z_0} & \frac{\partial\alpha_1}{\partial \dot{x}_0} & \frac{\partial\alpha_1}{\partial \dot{y}_0} & \frac{\partial\alpha_1}{\partial \dot{z}_0}\\\frac{\partial\alpha_2}{\partial x_0} & \frac{\partial\alpha_2}{\partial y_0} & \frac{\partial\alpha_2}{\partial z_0} & \frac{\partial\alpha_2}{\partial \dot{x}_0} & \frac{\partial\alpha_2}{\partial \dot{y}_0} & \frac{\partial\alpha_2}{\partial \dot{z}_0}\\\frac{\partial\alpha_3}{\partial x_0} & \frac{\partial\alpha_3}{\partial y_0} & \frac{\partial\alpha_3}{\partial z_0} & \frac{\partial\alpha_3}{\partial \dot{x}_0} & \frac{\partial\alpha_3}{\partial \dot{y}_0} & \frac{\partial\alpha_3}{\partial \dot{z}_0}\\\frac{\partial\delta_1}{\partial x_0} & \frac{\partial\delta_1}{\partial y_0} & \frac{\partial\delta_1}{\partial z_0} & \frac{\partial\delta_1}{\partial \dot{x}_0} & \frac{\partial\delta_1}{\partial \dot{y}_0} & \frac{\partial\delta_1}{\partial \dot{z}_0}\\\frac{\partial\delta_2}{\partial x_0} & \frac{\partial\delta_2}{\partial y_0} & \frac{\partial\delta_2}{\partial z_0} & \frac{\partial\delta_2}{\partial \dot{x}_0} & \frac{\partial\delta_2}{\partial \dot{y}_0} & \frac{\partial\delta_2}{\partial \dot{z}_0}\\\frac{\partial\delta_3}{\partial x_0} & \frac{\partial\delta_3}{\partial y_0} & \frac{\partial\delta_3}{\partial z_0} & \frac{\partial\delta_3}{\partial \dot{x}_0} & \frac{\partial\delta_3}{\partial \dot{y}_0} & \frac{\partial\delta_3}{\partial \dot{z}_0}\end{bmatrix} \cdot \begin{bmatrix}\Delta x_0\\\Delta y_0\\\Delta z_0\\\Delta \dot{x}_0\\\Delta \dot{y}_0\\\Delta \dot{z}_0\end{bmatrix}$$

By solving for the rightmost matrix, the corrected vector orbital elements can be found:

$$\vec{r} = \{x_0 + \Delta x_0, y_0 + \Delta y_0, z_0 + \Delta z_0\}$$
$$\dot{\vec{r}} = \{\dot{x}_0 + \Delta \dot{x}_0, \dot{y}_0 + \Delta \dot{y}_0, \dot{z}_0 + \Delta \dot{z}_0\}$$

Using the corrected vector orbital elements, the classical orbital elements were recalculated to a more accurate extent. Uncertainties for these elements were also determined based on the sample standard deviation of the LSPR residuals (with N – 3 degrees of freedom).

## III. DATA

The data from four observations of the asteroid were processed and reduced. The RA and Dec of the asteroid at each observation were calculated using several reference stars. The orbital elements of the asteroid were obtained using the RA and Dec of three of the observations.

### A. Asteroid Observations

The following images were taken using the 14" Meade telescope. Corrections had to be made to the time of observation recorded on the computer as it was 47 seconds behind actual UT. This difference was obtained after calling the US Naval Observatory's Master Clock.

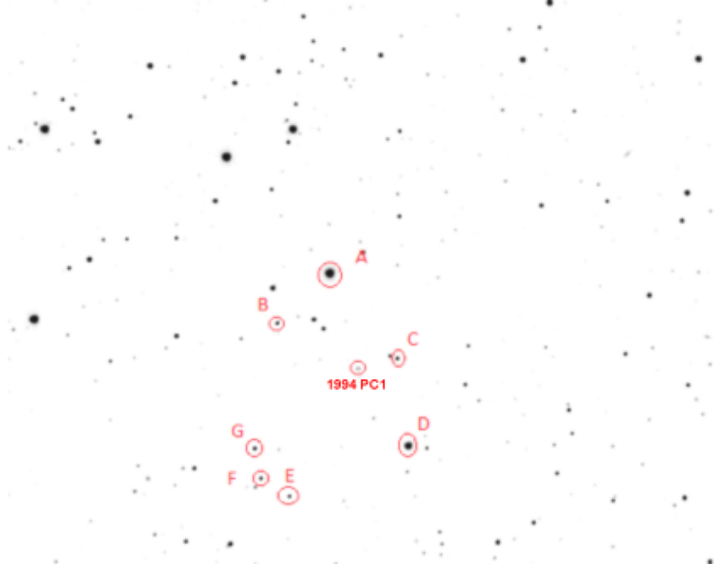

*Figure 3.1 Image of the asteroid taken on June 28th, 2011 at 05:13:21.296. Exposure length: 45 seconds. JD: 2455740.717608.*

*Table 3.2 List of the reference stars in Figure 3.1 and their residuals.*

| Letter on Chart and Star Name | X (pixel) | Y (pixel) | RA | DEC | Residuals |
|---|---|---|---|---|---|
| A: GSC 2588:2193 | 282.139 | 240.376 | 16h 44m 38.87s | +37° 03' 34.11" | RA: 0.0201 sec Dec: 0.0231" |
| B: GSC 2588:2260 | 236.421 | 284.235 | 16h 44m 30.53s | +37° 04' 48.71" | RA: -0.0150 sec Dec: 0.0199" |
| C: GSC 2588:2338 | 341.225 | 314.965 | 16h 44m 47.04s | +37° 06' 12.75" | RA: -0.01700 sec Dec: -0.0484" |
| D: GSC 2588:2192 | 350.760 | 391.267 | 16h 44m 47.15s | +37° 08' 43.33" | RA: 0.0053 sec Dec: 0.01978" |
| E: NOMAD | 246.927 | 435.799 | 16h 44m 29.36s | +37° 09' 45.75" | RA: 0.0172 sec Dec: 0.0500" |
| F: GSC 2588:2153 | 222.520 | 420.005 | 16h 44m 25.64s | +37° 09' 09.32" | RA: -0.0236 sec Dec: -0.0056" |
| G: GSC 2588:2235 | 216.823 | 393.866 | 16h 44m 25.25s | +37° 08' 17.14" | RA: 0.0130 sec Dec: -0.0588" |

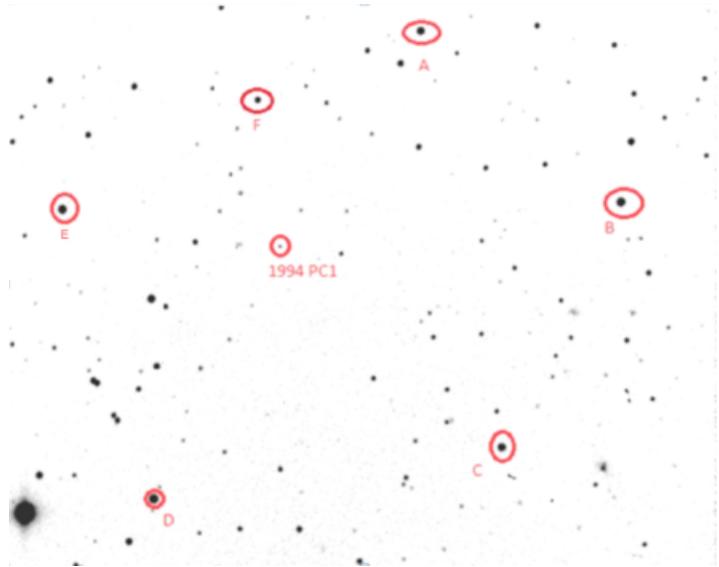

*Figure 3.3 Image of the asteroid taken on July 3, 2011 at 5:27:34.203. Exposure length: 45 seconds. JD: 2455745.727479*

*Table 3.4 List of the reference stars in Figure 3.3 and their residuals*

| Letter on Chart and Star Name | X (pixel) | Y (pixel) | RA | DEC | Residuals |
|---|---|---|---|---|---|
| A: UCAC3 241:121203 | 322.319 | 84.156 | 16h 39m 35.32s | +30° 02' 35.30" | RA: -0.0108 sec Dec: 0.0798" |
| B: UCAC3 240:122430 | 478.309 | 218.079 | 16h 39m 14.33s | +29° 57' 38.40" | RA: -0.0037 sec Dec: -0.2516" |
| C: UCAC3 240:122461 | 385.513 | 410.026 | 16h 39m 31.65s | +29° 51' 46.66" | RA: 0.0176 sec Dec: 0.2239" |
| D: UCAC3 240:122521 | 114.338 | 450.466 | 16h 40m 12.95s | +29° 51' 30.07" | RA: -0.0209 sec Dec: -0.0975" |
| E: UCAC3 240:122539 | 43.221 | 223.840 | 16h 40m 19.63s | +29° 59' 07.55" | RA: 0.0171 sec Dec: -0.0696" |
| F: UCAC3 241:121229 | 195.404 | 138.176 | 16h 39m 55.30s | +30° 01' 19.42" | RA: 0.0008 sec Dec: 0.1151" |

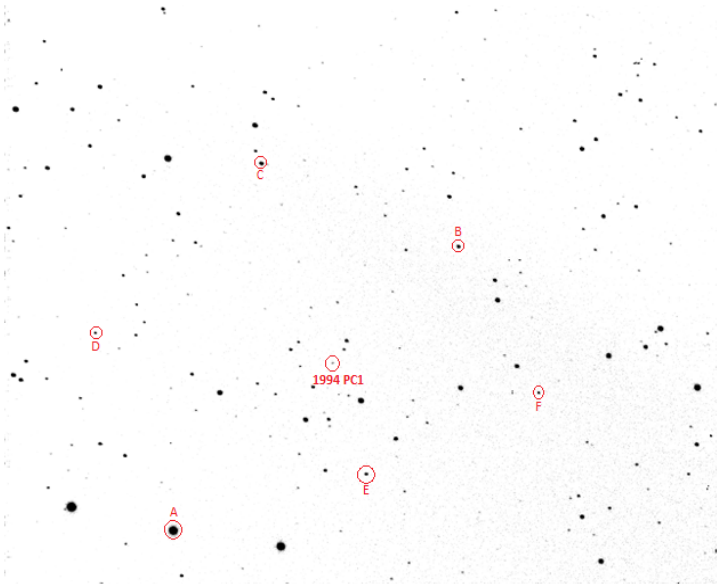

*Figure 3.5 Image of the asteroid taken on July 13, 2011 at 4:36:57.218. Exposure length: 37 seconds. JD: 2455755.692329*

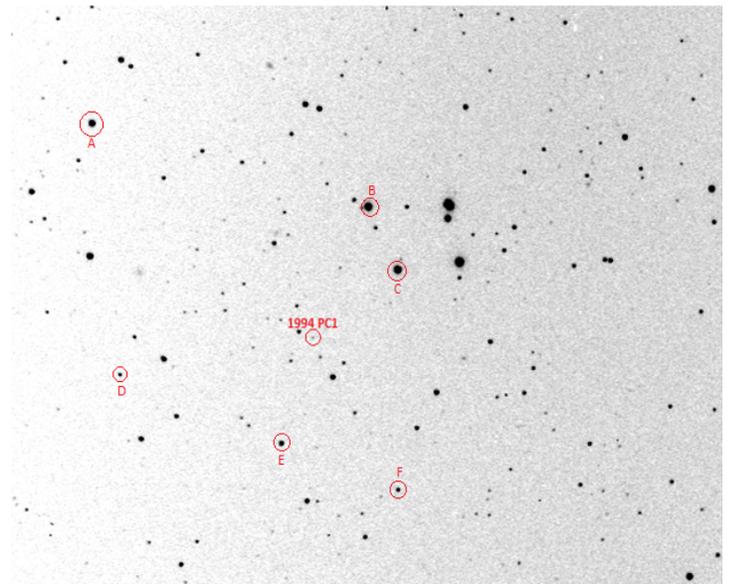

*Figure 3.7 Image of the asteroid taken on July 15, 2011 at 05:04:38.484. Exposure length: 37 seconds. JD: 2455757.711557*

*Table 3.6 List of the reference stars in Figure 3.5 and their residuals.*

| Letter on Chart and Star Name | X (pixel) | Y (pixel) | RA | DEC | Residuals |
|---|---|---|---|---|---|
| A: Tycho 1519:1427 | 151.485 | 464.494 | 16h 35m 15.554s | +15° 55' 46.867" | RA: -0.0095 sec Dec: -0.2912" |
| B: UCAC3 212:132005 | 407.586 | 213.056 | 16h 35m 46.16s | +15° 46' 40.40" | RA: -0.0053 sec Dec: -0.0599" |
| C: UCAC3 212:131946 | 230.658 | 139.539 | 16h 35m 21.16s | +15° 44' 57.70" | RA: 0.0038 sec Dec: -0.1103" |
| D: UCAC3 212:131908 | 81.373 | 289.449 | 16h 35m 03.37s | +15° 50' 23.21" | RA: -0.0003 sec Dec: 0.2258" |
| E: UCAC3 212:131981 | 324.731 | 414.442 | 16h 35m 38.16s | +15° 53' 30.80" | RA: 0.0172 sec Dec: 0.2415" |
| F: UCAC3 212:132024 | 479.636 | 342.470 | 16h 35m 57.90s | +15° 50' 35.53" | RA: -0.0059 sec Dec: -0.0059" |

*Table 3.8 List of the reference stars in Figure 3.7 and their residuals.*

| Letter on Chart and Star Name | X (pixel) | Y (pixel) | RA | DEC | Residuals |
|---|---|---|---|---|---|
| A: UCAC3 207:134628 | 79.880 | 123.425 | 16h 34m 53.59s | +13° 02' 54.54" | RA: 0.0081 sec Dec: 0.1118" |
| B: UCAC3 207:134703 | 324.939 | 192.767 | 16h 35m 27.30s | +13° 04' 13.88" | RA: -0.0090 sec Dec: -0.0435" |
| C: UCAC3 207:134714 | 351.054 | 245.065 | 16h 35m 31.60s | +13° 05' 49.71" | RA: 0.0023 sec Dec: -0.0680" |
| D: UCAC3 207:134636 | 104.677 | 332.555 | 16h 35m 00.11s | +13° 09' 35.61" | RA: -0.0107 sec Dec: -0.1347" |
| E: UCAC3 207:134683 | 247.859 | 389.664 | 16h 35m 20.09s | +13° 10' 54.52" | RA: 0.0067 sec Dec: 0.0392" |
| F: UCAC3 207:134717 | 351.331 | 428.218 | 16h 35m 34.47s | +13° 11' 46.19" | RA: 0.0026 sec Dec: 0.0953" |

## B. RA and Dec of Asteroid

*Table 3.9 Observation Data*

| Date (UT) | Julian Day | J2000 RA | J2000 RA (decimal hours) | J2000 Dec | J2000 Dec (decimal degrees) |
|---|---|---|---|---|---|
| June 28, 2011 05:13:21 | 2455740.71761 | 16h 44m 41.539s | 16.7449 | 37º 06' 20.29" | 37.1056 |
| July 3, 2011 05:27:34 | 2455745.72748 | 16h 39m 54.689s | 16.6652 | 29º 57' 31.52" | 29.9588 |
| July 13, 2011 04:36:57 | 2455755.69233 | 16h 35m 32.543s | 16.5924 | 15º 50' 26.58" | 15.8407 |
| July 15, 2011 05:04:38 | 2455757.71156 | 16h 35m 22.427s | 16.5896 | 13º 07' 56.92" | 13.1325 |

In order to find the orbital elements that most closely matched those on JPL, it was necessary to try all possible combinations of the four observations that were made. In the end, the elements for observations 1, 2, and 3 were closest to those reported by JPL for asteroid 1994 PC1. Refer to the appendix for the classical elements for all possible combinations of observations.

N.B. Due to differential corrections, which depend on various partial derivatives and ephemeris-generated RAs and Decs, the positive and negative deviations in Table 3.10 do not have the same absolute value.

*Table 3.11 Comparisons with JPL[4]*

| Quantity | Calculated Values | JPL Values[4] | Percent Difference |
|---|---|---|---|
| Semi-major axis (AU) | 1.3401 | 1.3462 | -0.453% |
| Eccentricity | 0.3272 | 0.3284 | -0.365% |
| Inclination (degrees) | 33.30 | 33.49 | -0.567% |
| Longitude of Ascending Node (degrees) | 118.01 | 117.91 | 0.085% |
| Argument of Perihelion (degrees) | 47.14 | 47.61 | -0.987% |
| Mean Anomaly for Middle Observation (degrees) | 65.35 | 64.76 | 0.911% |
| Time of Last Perihelion (JD) | 2455642.88 | 2455643.10 | -8.96 x $10^{-6}$ % |

## C. Orbital Elements

*Table 3.10 Classical Orbital Elements*

| Element and Units | Value | Uncertainty (Due to LSPR Residuals) | Uncertainty (Assuming 1" Residuals) |
|---|---|---|---|
| Semi-major Axis (AU) | 1.3401 | +0.0001 AU or -0.0004 AU | +0.0001 AU or -0.0005 AU |
| Eccentricity | 0.3272 | +9e-05 or -0.0005 | +6e-05 or -0.0006 |
| Inclination (degrees) | 33.30 | +0.05 degrees or -0.01 degrees | +0.05 degrees or -0.002 degrees |
| Longitude of Ascending Node (degrees) | 118.01 | +0.09 degrees or -0.02 degrees | +0.08 degrees or -0.04 degrees |
| Argument of Perihelion (degrees) | 47.14 | +0.005 degrees or -0.08 degrees | +0.006 degrees or -0.05 degrees |
| Mean anomaly (degrees) for middle observation | 65.35 | +0.03 degrees or -0.02 degrees | +0.05 degrees or -0.03 degrees |
| Time of Last Perihelion (JD) | 2455642.88 | +0.007 days or -0.03 days | +0.008 days or -0.04 days |

## D. Ephemeris Check

*Table 3.12 Ephemeris-Generated and Observed Equatorial Coordinates.*

| Time (JD and UT) | RA Observed | | Dec Observed | | RA Calculated | | Dec Calculated | |
|---|---|---|---|---|---|---|---|---|
| JD 2455740.71761 June 28, 2011 05:13:21 | 16h 44m 41.539s | 16.7449 hrs | 37º 06' 20.29" | 37.1056º | 16h 44m 41.640s | 16.7449 hrs | 37º 04' 53.04" | 37.0814º |
| JD 2455745.72748 July 3, 2011 05:27:34 | 16h 39m 54.689s | 16.6652 hrs | 29º 57' 31.52" | 29.9588º | 16h 39m 54.720s | 16.6652 hrs | 29º 55' 52.32" | 29.9312º |
| JD 2455755.69233 July 13, 2011 04:36:57 | 16h 35m 32.543s | 16.5924 hrs | 15º 50' 26.58" | 15.8407º | 16h 35m 32.640s | 16.5924 hrs | 15º 48' 32.04" | 15.8089º |
| JD 2455757.71156 July 15, 2011 05:04:38 | 16h 35m 22.427s | 16.5896 hrs | 13º 07' 56.92" | 13.1325º | 16h 35m 22.560s | 16.5896 hrs | 13º 06' 01.08" | 13.1003º |

*Note that the values generated by the ephemeris program are very close to those observed.*

### E. Photometry

*Table 3.13. Photometry measurements*

| Reference Stars | V Magnitude of the Star | V Magnitude of the Asteroid |
|---|---|---|
| Tycho 968:625 | 11.11 | 17.645 |
| Tycho 168:7 | 10.91 | 17.687 |
| Tycho 968:669 | 11.12 | 17.465 |

*The mean value of the magnitude is 17.599. The standard deviation is 0.118. Therefore, taking into account significant digits, the magnitude of the asteroid is 17.6 $\pm$ 0.1*

During one of the observations, images of the asteroid were taken using a green visual filter. These images were then used to determine the absolute V magnitude of the asteroid. TheSkyX only had information on V magnitude for three of the stars in the field of view of the asteroid. As a result, the magnitudes of these stars were calibrated through MaxIm DL using the values from TheSkyX. Using these calibrated values, the magnitude of the asteroid was found to be 17.6 $\pm$ 0.1. There were three slightly varying results; therefore, the mean value and standard deviation had to be taken.

### F. Previously Accepted Values

*Table 3.14. JPL Horizons' generated values for July 2$^{nd}$, 2011*

| Orbital Element | JPL Horizons Value |
|---|---|
| Eccentricity | 0.328352 |
| Semi major Axis (AU) | 1.346226 |
| Inclination (degrees) | 33.48687 |
| Longitude of Ascending Node (degrees) | 117.9127 |
| Argument of Perihelion (degrees) | 47.6111 |
| Mean anomaly (degrees) | 64.75986 |
| Time of last Perihelion (JD) | 2455643.09636 |

*Note: JPL Horizons does not give uncertainties for its generated values.*

### G. Simulation

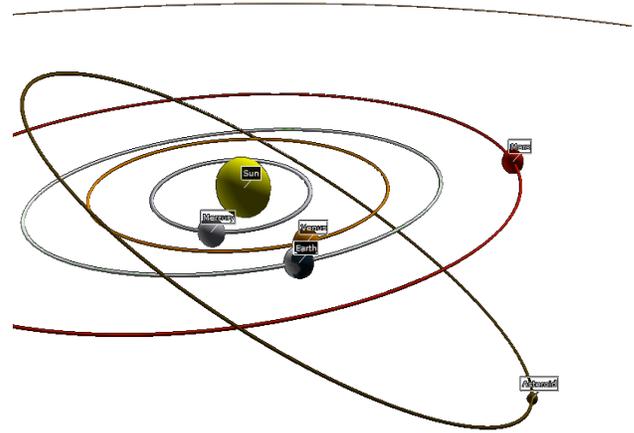

*Figure 3.15. Simulation of the asteroid in a realistic Solar System to show gravitational perturbations. Sizes of Sun and planets are not to scale. Distances, inclination angles, and shapes of orbits are to scale.*

Based on a computer simulation, it was determined that the point of closest approach to Earth in the next 100 years is $\approx 7.396 \times 10^9$ m or $\approx 0.0494$ AU. The time of closest approach shall be around January 21, 2056 (JD 2472018). Note that this is highly dependent on the accuracy of our results. Fortunately, due to Verlet integration, cumulative error in position could be minimized (to around $O(\Delta t)^4$). However, due to chaos theory and potential inaccuracies in orbital elements, these measurements may not be as accurate in the distant future.

### H. Visualization

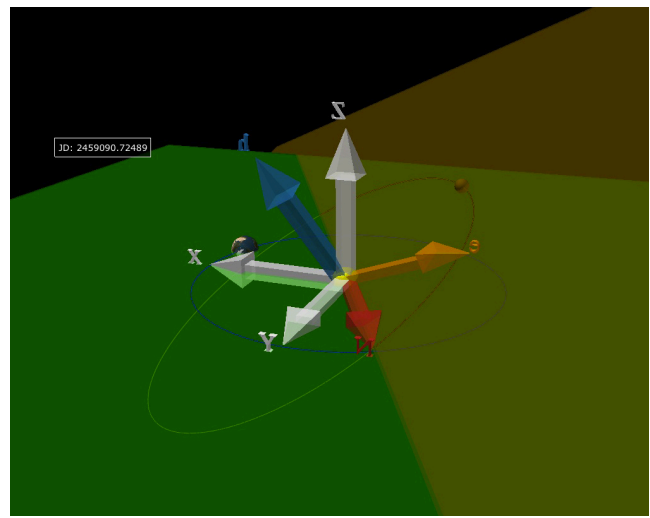

*Figure 3.16. A VPython visualization of the asteroid's orbital element vectors and its orbit around the Sun. Angular momentum vector is in blue, longitude of ascending node is in red, and eccentricity (pointing to perihelion) is in orange. Asteroid is in brown. Earth's (almost) circular orbit (blue) in the plane of the ecliptic (green) is shown for scale of distances (which are to scale). Sizes (i.e. radii) of Earth, Sun, and asteroid are not to scale.*

## IV. Preliminary Interpretation of Results

The visualization (for July 2011) shows that the asteroid is near opposition and moving downwards from above the plane of the ecliptic. This matches with the JPL[4] ephemeris predictions, which predicted that the asteroid goes from retrograde to prograde during our observation schedule. Furthermore, our results corroborate the fact that the asteroid is a potentially hazardous object, since it reaches very close to the Earth during its orbit.

## V. Conclusion

According to the data used in the report, collected in July 2011, the Near-Earth Asteroid 1994 PC1 has an eccentricity of 0.3272 (+0.00009 or –0.0005), a semi major axis of 1.3401 AU (+0.0001 AU or -0.0004 AU), an inclination angle of $33.30°$ (+$0.05°$ or -$0.01°$), a longitude of the ascending node of $118.01°$ (+$0.09°$ or -$0.02°$), an argument of perihelion of $47.14°$ (+$0.005°$ or -$0.08°$), and a time of last perihelion of JD 2455642.88 (+0.007 or -0.03). The asteroid was also found to have an absolute magnitude of $17.599 \pm 0.118$.

These data are not completely consistent with JPL's current information concerning 1994 PC1. The statistical uncertainty derived from the Least Squares Plate Reduction Method residuals does not succeed in accounting for these differences. The LSPR method of recalculating the asteroid's Right Ascension and Declination is a possible source of systematic error: the reference stars were picked according to the lowest associated residuals, and were not all equidistant from (and circularly distributed around) the asteroid. Compared with the 468 1994 PC1 observations in the JPL database, the research conclusions presented in this report are relatively reliable. Extrapolating from the set of calculated orbital elements, the asteroid's ephemeris was found and compared to that proposed by NASA[4]. The difference in equatorial coordinates was minimal for the computed Julian Date and our data could (potentially) reveal new information concerning 1994 PC1.

The ephemeris-generated values for all four observations were very accurate (refer to Table 3.12). A visualization of the Earth's and 1994 PC1's orbits showed that the asteroid will most likely not collide with our planet in the near future, though the point of closest approach will be January 21, 2056. This was determined through a 100-year-long simulation of the inner Solar System and the asteroid (using Verlet integration). In order to reach a more definite conclusion on possible asteroid impact, however, more data would be required. Another method of orbit determination (such as Laplace's technique) could be used as well to confirm the results received through the Method of Gauss. Furthermore, a telescope with a better angular resolution (i.e. larger aperture) and adaptive optics would help not only correct blurry images (due to poor seeing), but also allow for more accurate centroiding of reference stars and our faint asteroid. This in turn would help determine more accurate RA and Dec values through LSPR.

## VI. Acknowledgements

APPENDIX

*Table A1: Classical Orbital Elements for Observations 1, 2, and 3*

| Element and Units | Value | Uncertainty (With LSPR Residuals) | Uncertainty (Assuming 1 arcsecond residuals) |
|---|---|---|---|
| Semi-major Axis (AU) | 1.3401 | +0.0001 AU or -0.0004 AU | +0.0001 AU or -0.0005 AU |
| Eccentricity | 0.3272 | +9e-05 or -0.0005 | +6e-05 or -0.0006 |
| Inclination (degrees) | 33.30 | +0.05 degrees or -0.01 degrees | +0.05 degrees or -0.002 degrees |
| Longitude of Ascending Node (degrees) | 118.01 | +0.09 degrees or -0.02 degrees | +0.08 degrees or -0.04 degrees |
| Argument of Perihelion (degrees) | 47.14 | +0.005 degrees or -0.08 degrees | +0.006 degrees or -0.05 degrees |
| Mean anomaly (degrees) for middle observation | 65.35 | +0.03 degrees or -0.02 degrees | +0.05 degrees or -0.03 degrees |
| Time of Last Perihelion (JD) | 2455642.88 | +0.007 days or -0.03 days | +0.008 days or -0.04 days |

*Table A2: Classical Orbital Elements for Observations 1, 2, and 4*

| Element and Units | Value | Uncertainty (With LSPR Residuals) | Uncertainty (Assuming 1 arcsecond residuals) |
|---|---|---|---|
| Semi-major Axis (AU) | 1.3385 | +0.0002 AU or -6e-05 AU | +0.0002 AU or -8e-05 AU |
| Eccentricity | 0.3266 | +0.0002 or -2e-05 | 0.0003 or -3e-05 |
| Inclination (degrees) | 33.27 | +0.0009 degrees or -0.01 degrees | +0.003 degrees or -0.009 degrees |
| Longitude of Ascending Node (degrees) | 118.06 | +0.004 degrees or -0.02 degrees | +0.004 degrees or -0.02 degrees |
| Argument of Perihelion (degrees) | 47.00 | +0.02 degrees or -0.006 degrees | +0.03 degrees or -0.005 degrees |
| Mean anomaly (degrees) for middle observation | 65.50 | +0.01 degrees or -0.01 degrees | +0.01 degrees or -0.01 degrees |
| Time of Last Perihelion (JD) | 2455642.81 | +0.01 days or -0.001 days | +0.02 days or -0.003 days |

*Table A3: Classical Orbital Elements for Observations 1, 3, and 4*

| Element and Units | Value | Uncertainty (With LSPR Residuals) | Uncertainty (Assuming 1 arcsecond residuals) |
|---|---|---|---|
| Semi-major Axis (AU) | 1.3330 | +6e-05 AU or -0.0003 AU | +1e-04 AU or -0.0004 AU |
| Eccentricity | 0.3254 | +7e-05 or -0.0003 | +1e-04 or -0.0003 |
| Inclination (degrees) | 33.07 | +0.03 degrees or -0.004 degrees | +0.03 degrees or -0.005 degrees |
| Longitude of Ascending Node (degrees) | 118.11 | +0.06 degrees or -0.01 degrees | +0.06 degrees or -0.01 degrees |
| Argument of Perihelion (degrees) | 46.61 | +0.01 degrees or -0.08 degrees | +0.01 degrees or -0.09 degrees |
| Mean anomaly (degrees) for middle observation | 72.46 | +0.01 degrees or -0.02 degrees | +0.02 degrees or -0.02 degrees |
| Time of Last Perihelion (JD) | 2455642.55 | +0.005 days or -0.02 days | +0.006 days or -0.03 days |

*Table A4: Classical Orbital Elements for Observations 2, 3, and 4*

| Element and Units | Value | Uncertainty (With LSPR Residuals) | Uncertainty (Assuming 1 arcsecond residuals) |
|---|---|---|---|
| Semi-major Axis (AU) | 1.3399 | +4e-05 AU or -0.0002 AU | +4e-05 AU or -0.0002 AU |
| Eccentricity | 0.3270 | +3e-05 or -0.0002 | +7e-05 or -0.0002 |
| Inclination (degrees) | 33.28 | +0.02 degrees or -0.006 degrees | +0.02 degrees or -0.007 degrees |
| Longitude of Ascending Node (degrees) | 117.99 | +0.03 degrees or -0.01 degrees | +0.04 degrees or -0.02 degrees |
| Argument of Perihelion (degrees) | 47.14 | +0.01 degrees or -0.03 degrees | +0.02 degrees or -0.04 degrees |
| Mean anomaly (degrees) for middle observation | 71.71 | +0.01 degrees or -0.007 degrees | +0.01 degrees or -0.01 degrees |
| Time of Last Perihelion (JD) | 2455642.84 | +0.004 days or -0.02 days | +0.005 days or -0.02 days |